# The influence of the stationary electric field upon the resonance absorption of microwaves in HeII

A.S. Rybalko, S.P. Rubets, E.Ya. Rudavskii, V.A. Tikhiy,
R. Golovachenko*, V.N. Derkach*, S.I. Tarapov*

B.I.Verkin Institute for Low Temperature Physics and Engineering,
National Academy of Sciences of Ukraine,
47, Lenin Ave., Kharkov, 61103.
* Institute for Radiophysics and Electronics,
National Academy of Sciences of Ukraine,
12, St. Proskury, Kharkov, 61085.
rybalko@ilt.kharkov.ua

## Abstract

The interaction between electromagnetic microwaves (40-200 GHz) and superfluid helium in a stationary electric field has been investigated experimentally. It is found that the narrow line of resonance absorption at the roton frequency is split in the electric field into two symmetric lines. The splitting magnitude increases almost linearly with the electric field, which suggests a linear Stark effect. The results obtained point of orientational polarizability and dipole moment (~$10^{-34}$ C·m) in HeII. It is shown that the spectral line profile consists of two parts – a narrow line of resonance absorption (or induced radiation when superfluid stream are generated) and a broad "background". The "background" with agrees well with the latest neutron data for the roton line.

## 1. Introduction

The He atom is a spherically symmetric system with no permanent dipole moment. This may be the reason for the modest interest in the electric properties of liquid helium. The electric activity detected in HeII recently [1, 2] have been a rather surprising finding. The investigations of the interaction between electromagnetic microwaves and superfluid helium [3-5] furnished new information in which the main result was the resonance absorption of microwaves at the frequency corresponding to the lowest roton energy. The resonance absorption occurs in parallel with ordinary dielectric losses. As in neutron scattering experiments [6], the roton line was also observed above the λ-point, which is still to be explained accurately.

It appeared later that the detected effect were highly sensitive to directed superfluid flows that were generated in HeII with special thermal guns equipped with heaters. The flow rate was controlled by varying the power fed to the heater. The resonance line changed dramatically with in increasing flow rate: the absorption of microwave photons turned into their induced radiation [7]. The effects were explained qualitatively assuming that HeII is a two-level system in which the energy difference between the levels is equal to the energy of a roton.

The investigation of microwave-HeII interaction involves another yet unanswered question of a possible effect of a *dc* electric field on resonance absorption. In multilevel systems the electric field effect usually makes itself evident in shifting and splitting resonance lines (Stark effect).

The new experiments reported here are attempts to answer the question whether this effect exists in HeII.

## 2. Experimental technique

The microwave technique used in these experiments is basically similar to that in [3-5]. There are however some distinctions. Firstly, the distance between the adjacent whispering gallery modes was reduced. For this purpose a quartz resonator was replaced with dielectric leucosapphire disk resonator 1 (Fig. 1) because the dielectric permittivity of leucosapphire is several times higher than that of quartz. We could thus increase considerably the number of observable azimuthal whispering gallery modes and, hence, the number of experimental points in the frequency range (40-200 GHz) of the experiment. A resonator 20 mm in diameter immersed in liquid helium was the main measuring element operating at the frequency of the whispering gallery mode. Exciting and receiving waveguide – antennas 2 were fixed in the plane of resonator.

Secondly, an orifice 3 mm in diameter was made at the center of the resonator to house metallic electrode 3. A constant potential difference was applied between electrode 3 end cell frame 4. The applied voltage was 500-4000 V. In this arrangement the liquid near the cylindrical surface of the resonator experiences a *dc* electric field excited by applied voltage and *ac* electric field generated by a traveling microwave ($E_{dc}$ and $E_{ac}$, respectively). Both the fields ($E_{dc} >> E_{ac}$) are directed along the radius off the center.

Thermal guns 5 were used in our previous experiments. The spectrum of microwave absorption in HeII was measured at varying powers of the heater and *dc* electric fields.

## 3. Spectral line profile

The profiles of the spectral lines obtained on passing an electromagnetic wave (40-200 GHz) through liquid helium are similar in the experiments with leucosapphire and quartz [3, 4] resonators.

The typical resonance curves for one of the whispering gallery modes taken in vacuum and superfluid helium are shown in Figs. 2a, b, respectively. It is seen that the Q-factor and the signal amplitude are smaller in the liquid due to dielectric losses. Besides, the resonance line has a feature at a certain temperature-dependent frequency: it shows up as a very narrow line-dip when the thermal gun is off (Fig. 2b) and changes drastically into a very narrow line-spike when the gun is on. The effect is explained qualitatively within the model for a two-level system assuming resonance absorption of microwave photons when the gun is off and their induced radiation under the heat-flow condition.

Note that the spectral line profile, when plotted, has two distinct parts – a narrow line of resonance absorption (or induced radiation when the gun is on) and a "background" shaped as broad side wings (Fig. 3). The width of the resonance in the wings agrees well with the recent neutron scattering measurements of the roton line width [6, 7]. It is seen in Fig. 3 that the operating thermal gun shifts the "background" slightly towards lower frequencies. The frequency of the narrow line remains invariant and its amplitude is dependent on the power $Q$ fed to the thermal gun. As $Q$ grows, absorption of microwave photons changes into induced radiation.

The narrow line in the "background" wings is rather unexpected because nothing of this kind had been observed either in neutron [7] or in Raman [8, 9] scattering experiments. The reason may be that the instrumental width of the radiation from the generator on the backward-wave tube restricted in these experiments by a high-Q whispering gallery mode. Our value of the width is about $5 \cdot 10^{-6}$ K, i.e. four of five orders of magnitude lower than in [7-9] which makes it possible to observe this narrow line in the spectrum. We are unaware of any other observation of such a line in spectra of condensed systems. The only exception is the Mössbouer effect in which the spectrum has narrow line against the phonon “background”.

## 4. Resonance line splitting in an electric field

The evolution of resonance line referring to the microwave photon absorption at the roton frequency in an increasing dc electric field $E_{dc}$ is illustrated in Fig. 4. At $E_{dc}$ = 0 (thermal gun is off) the dip (a very narrow roton line) coincides with the azimuthal mode maximum $m$ = 128 (Fig. 4a). We close a moderately low temperature to obtain a resonance line no wider than 100 kHz at the bottom [3]. On applying an electric field, the line is split into two symmetrical parts (Fig. 4b). Note that the resonance lines in Fig. 4 were obtained by superimposing 30 frequency scans to average the spread of experimental results. Temperature variations affected only the resonance frequency (due to a change in the lowest energy of the roton) and the Q-factor of the line, which is in full accord with [3].

The magnitude of the resonance level splitting $\Delta f$ increases almost linearly with the electric field $E_{dc}$ (Fig. 5). These experimental dependences bear clear evidence of the linear Stark effect. In accordance with the ordinary quantum-mechanical interpretation of the Stark effect, it is possible to assume that a helium atom (or any other quantum system) with the energy $E$ in a zero electric field gains additional energy ΔE when the external field is applied. This is because the electron shell of the atom becomes polarized and an induced dipole moment appears. Thus, the electric field splits the degenerate level into two Stark sublevels $E+\Delta E$ and $E-\Delta E$. It is the Stark splitting, rather than the shift of the level that points to the presence of a dipole moment in the system [10].

The electric field dependence of the resonance frequency allowing for the splitting $\Delta f$ (Fig. 5) can be approximated by the linear dependence

$$f = f_0 + kE_{dc}, \tag{1}$$

where $f_0 = 1.803\cdot10^{11}$ Hz, $k$ = 42.65 Hz cm/kV.

The additional energy of the system with the electric dipole moment $\vec{P}$ in the electric field $\vec{E}$ is

$$W = -(\vec{P}\vec{E}), \tag{2}$$

On the other hand, the energy $W$ can be expressed in terms of splitting

$$W = h\Delta f, \tag{3}$$

We can thus estimate the dipole moment of the system in the electric field $E$:

$$P = \frac{h\Delta f}{E} \cong 3\cdot10^{-34} \tag{4}$$

It is not yet clear what quantum system this dipole moment refers to.

Normally, the dipole moment is associated with polarizability of atoms, i.e. deformability of their electron shells in the electric field $E$. In this case the atoms gains an induced dipole moment

$$\mathbf{p}_{ind}=\alpha\mathbf{E} \quad (5)$$

where $\alpha$ is the atomic polarizability. For helium at $T$ = 1,5 K $\alpha$ = 0.1232 cm$^3$/mol = 2.1·10$^{-11}$ $Am^2s/B$ [10]. The potential energy of such induced dipole in the homogeneous field $E$ = 4·10$^5$ V/m is

$$W= \alpha E^2/2=1,6\cdot 10^{-30} \text{ W*s} \quad (6)$$

In this case the shift of the spectral line frequency could be $\delta f = W/h$ = 2.55 kHz, which is almost two orders of magnitude smaller than the shift in the experiment (340 kHz). Hence, the induced polarization in our experiment (the formation of an induced dipole in the system) is inadequate to explain the effect observed.

Note that E in Eq. (5) stands for the local electric field at the atom site. $E = E_{dc}$ is possible only for an isolated atom. In liquid helium the atoms experience an additional internal field $E_{inetrn}$ excited by the surrounding dipoles. $E_{inetrn}$

Is much larger/stronger (see the above estimate) than the applied field. Extrapolating the data of Fig. 3 to zero frequency, we obtain $E_{inetrn} \sim 10^{12}$ V/m, which agrees with the Coulomb field excited by the nucleus on the 2S-state electron orbit.

The dipole moment can be estimated from the available experimental data on the permittivity of HeII [12] and the temperature dependence of its density [11]. Using the Clausius-Mossotti equations we obtain reliable information about the polarizability of liquid helium as a function of density and temperature. The temperature dependence of polarizability of superfluid helium $\alpha(T)$ can be treated as a sum of a constant quantity $\alpha_0$ and a small temperature-dependent addition $\alpha_1(T)$ which increases at decreasing temperature [11]:

$$\acute{\alpha}(T) =\alpha_0 +\alpha_1(T) \quad (7)$$

Eq. (7) resembles the known Langevin-Debye expression (see [12]) relating polarizability to the dipole moment of the particle. The Langevin-Debye equation was derived within the gas model of polar molecules. Its first term corresponds to temperature-independent inductive polarization caused by elastic shifts of atoms. The second term describes orientational polarization dependent on the thermal motion of atoms. In this model the second term of Eq. (7) had the form

$$\alpha_1(T)= P^2/3kT \quad (8)$$

$\alpha_1(T)$ obtained from the data of [11] is 2 10$^{-4}$ cm$^3$/mol or 3.3 10$^{-44}$ Am$^2$s/V in S/units. Then, the dipole moment calculated by Eq. (8) is P = 10$^{-33}$ C m, which is quite close to the value obtained in this study. This suggests that the concept of inherent nonpolarity of superfluid helium atoms is rather disputable.

The dipole moment can also be estimated from the results of the recent investigation of the second sound in HeII [1] in which an electric response to the second sound wave was detected and the potential difference $\delta U$ generated by the propagating wave in the second sound capacitor/resonator was measured. The capacitor plates were an electrode fixed at the site of a second sound detector and the resonator body. The spacing between the capacitor plates in the cell was $a$ = 1.05 mm. The potential difference at $T$ = 1.5 K was $\delta U$ = 4.3·10$^{-8}$ V, which corresponds to the charge $Q$ = 6.5·10$^{-15}$ C in the capacitor. We can thus anticipate a macroscopic dipole moment $P_{macr} = aQ$ = 6.5·10$^{-18}$ C·m in the system.

The electric charge on the capacitor plates implies polarization of He atoms. The number of atoms within the measuring 1.05 mm – size-cell is $n$ = 2.2 10$^{19}$. Let $\delta n$ of them be polarized by the second sound wave. We can assume $\delta n/n = \delta T/T$, where $\delta T$ is the change in the temperature caused by the second sound wave. Then, $\delta n = 1.5\cdot10^{16}$ atoms at $T = 1.5$ K and $\delta T = 10^{-3}$ K. Since the macroscopic dipole moment consists of the microscopic dipole moments of individual atoms, the dipole moment of a single atom is $P = P_{macr}/\delta n = 3\cdot10^{-34}$ C·m, which practically coincides with the conclusion of this study. The macroscopic dipole moment $P_{macr} = 2.8\cdot10^{-34}$ C·m found from the processed results for two (1.05 mm and 28.0 mm long) second-sound resonators is constant within ± 30 % and does not depend on temperature.

Assuming that the dipole and electron charges are nearly equal, the dipole arm is $1.75\cdot10^{-15}$ mm. According to Eq. (5), this shift can be obtained by applying the external electric field $E_{dc} \sim 10^7$ V/m, which is on the verge of electrical breakdown in liquid helium. It is likely that such a shift can be caused by the internal electric field excited by the interaction forces in the quantum system. Note that the distance $\sim 10^{-15}$ m is shorter than the atomic dimensions and is not in use in hydrodynamics.

## 5. Conclusions

The Stark effect (spectral line splitting in an electric field) has been observed for the first time in superfluid helium. The Stark splitting was detected at the frequency corresponding to the lowest roton energy and its magnitude depended linearly on the applied electric field. Commonly, the Stark effect is associated with polarization of the system and the appearing (or existing) electric dipole moment. Such dipole moment was estimated ($3\cdot10^{-34}$ C·m) through analyzing and comparing data of other experiments. It is assumed that this is the dipole moment of a helium atom. The value of the dipole moment obtained in our resonance experiment is verified by the analysis of nonresonance measurements of the temperature dependence of liquid helium polarizability and the electric response in the second sound wave in HeII.

The results obtained entail new questions and problems open to discussion. The Landau model treats the properties of superfluid helium in terms of elementary excitations-phonons and rotons rather than individual atoms. The correspondence between the spectral line and the roton frequency and between the dipole moment and the atom calls for a thorough analysis of the atomic interaction in HeII. It is important to identify the quantum system in superfluid helium which is responsible for the Stark effect. It is also good to ascertain the origin of the internal electric field in helium and to answer the question whether helium atoms can be polar.

Another range of questions are concerned with the mechanism if inducing a dipole moment in helium atoms. The dipole moment is induced in an atom when its spherical symmetry is disturbed. It is likely that the dipole moments of the elementary particles making up an atom become essential in this case. The question whether such elementary particles can have a dipole moment has been much discussed lately [14-16]. According to quantum mechanics, a charged elementary particle can have a dipole moment if its centre of charge does not coincide with its center of mass. The electric dipole moment of an electron together with the hyperfine interaction can induce an electric dipole moment of a closed-shell atom (see [17]). It is not yet clear whether this speculation is applicable to a helium atom.

## Acknowledgement

The study was supported with CRDF (Project 2853) and STCU (Project 3718) grants.

**Figure Captions**

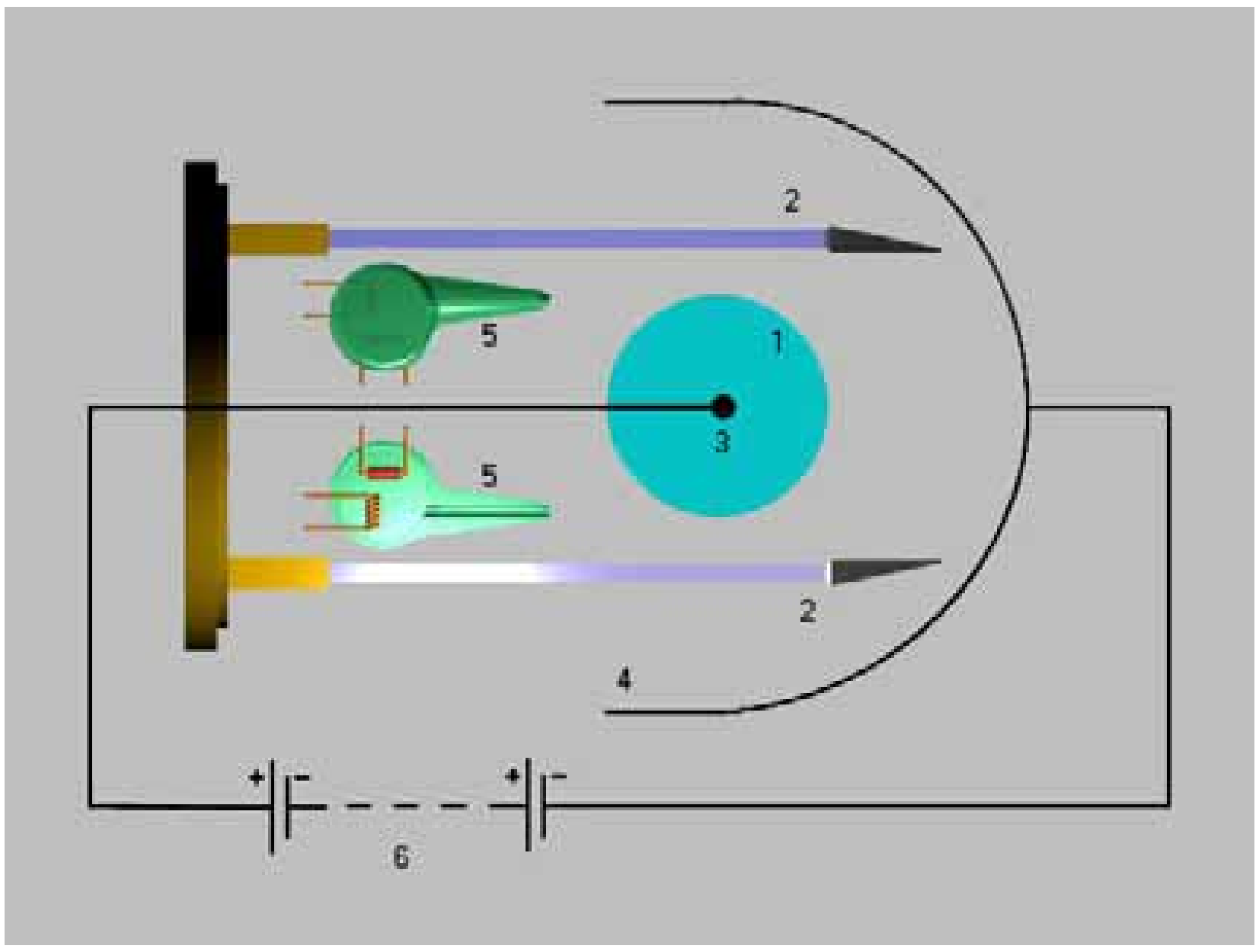

Fig. 1. Basic diagram of measurement in external *dc* electric field.
1 – dielectric disk resonator; 2 – dielectric antennas, 3 – metallic electrode, 4 – part of the metallic body of the cell with HeII, 5 – thermal guns, 6 – d.c. voltage source.

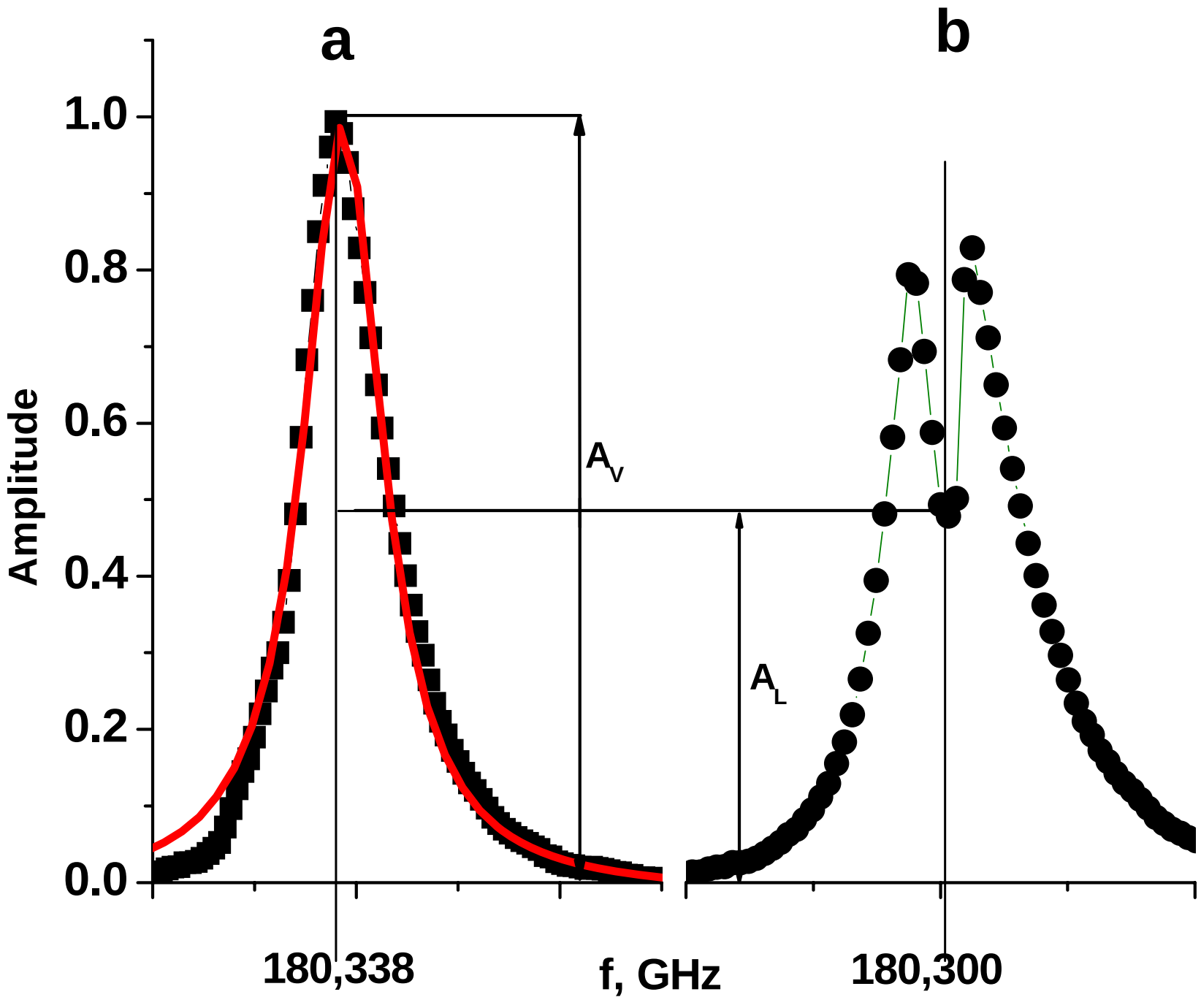

Fig. 2. Resonance curves of whispering gallery modes of a leucosapphire resonator: a – vacuum, mode $m$ = 128, $T$ = 1.4 K; b – HeII, mode $m$ = 128, $T$ = 1.4 K

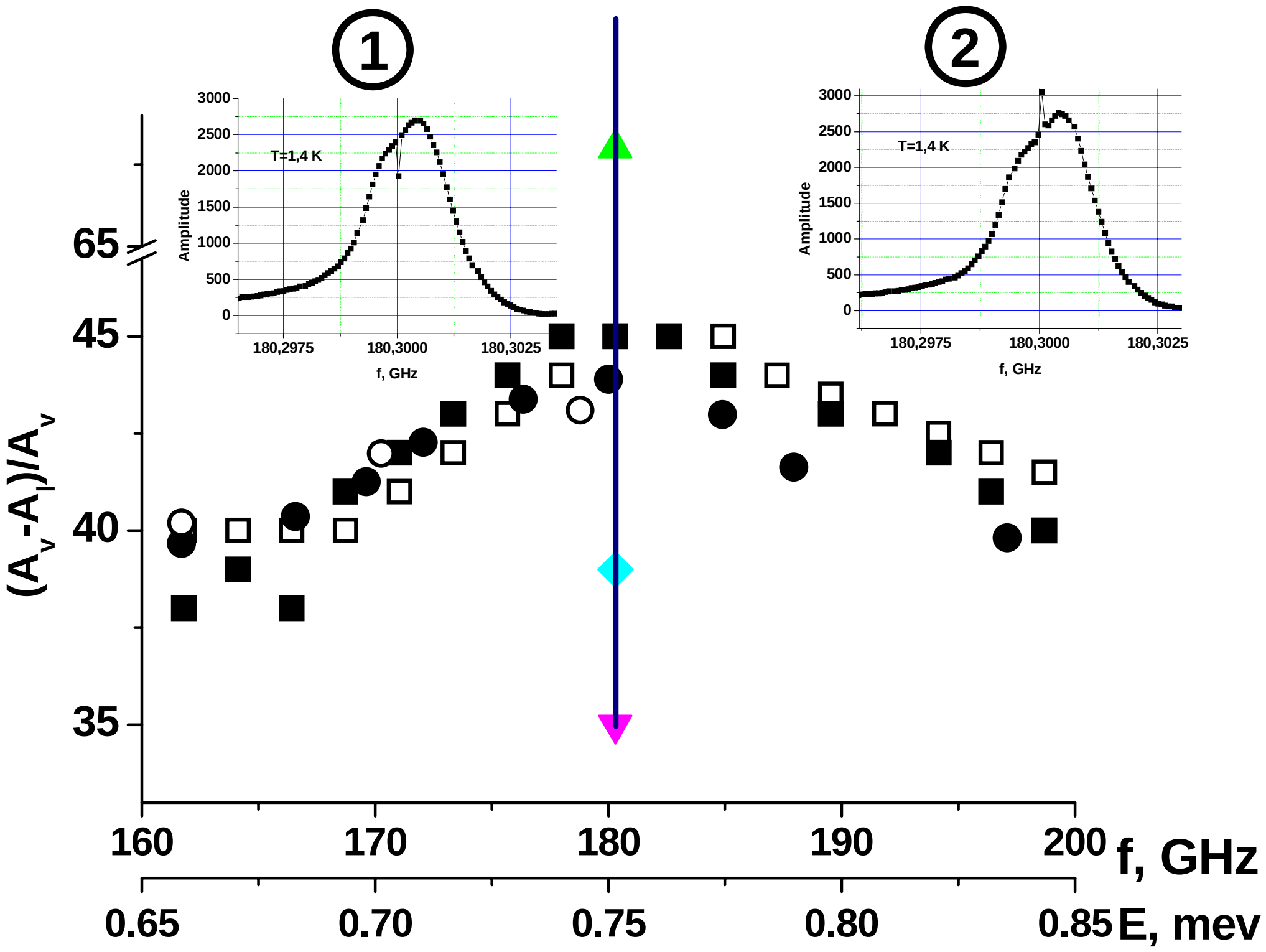

Fig. 3. Spectral line profile ($m$ = 128, $T$ = 1.4 K) and its comparison with the roton line derived from neutron data [7]. Inset 1: resonance absorption of microwave photons. Inset 2: induced photon radiation. This study: ■, □ – thermal gun is on (7 W/cm$^2$) and off (7 W/cm$^2$), respectively. Neutron data [7]: ○, ● – $T$ = 1.34, 1.5 K, respectively, $P$ = 0.4 Bar. Points on the narrow resonance absorption line show the power of the heat flow from 0 (upper point) to 7 W/cm$^2$ (lower point).

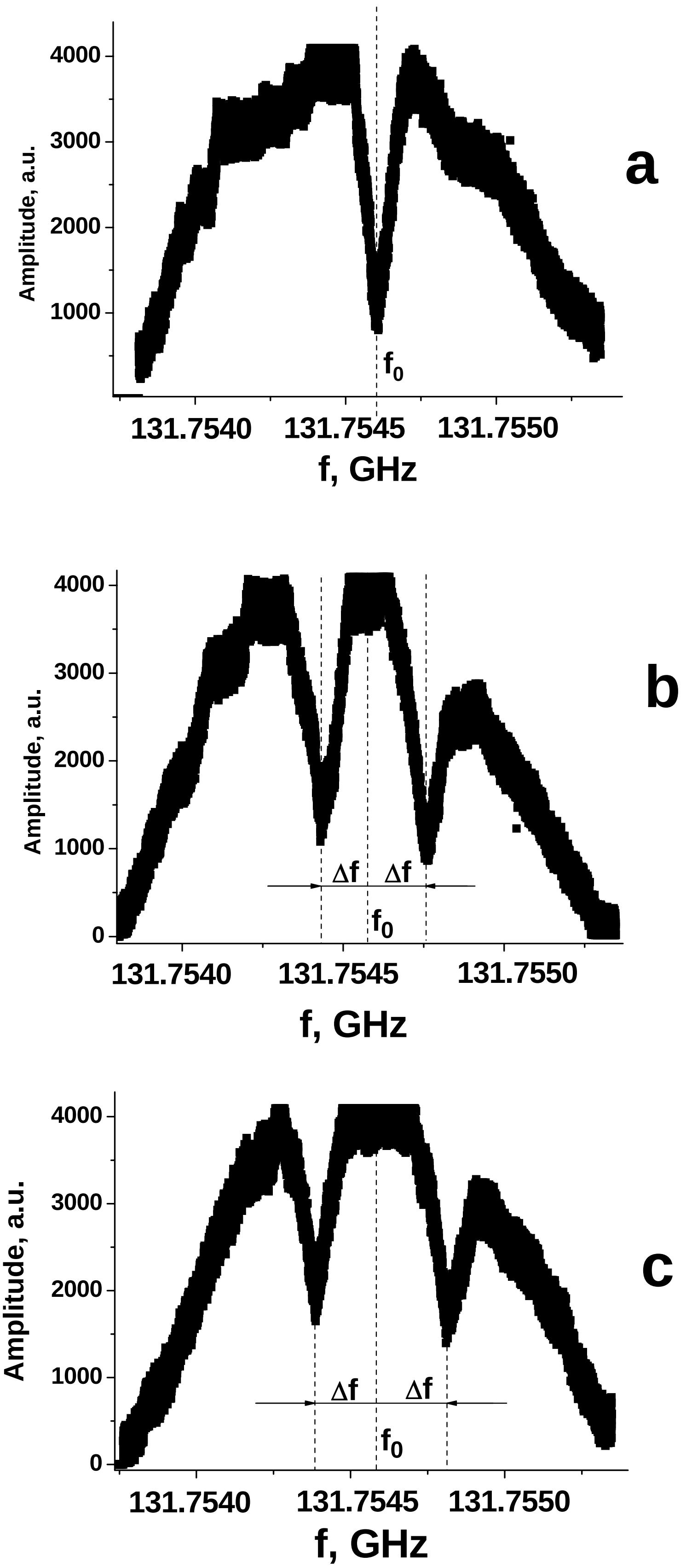

Fig. 4. Profile variations of the resonance microwave absorption line in electric field $E_{dc}$: a) $E_{dc} = 0$, b) $E_{dc} = 2$ kV/cm, c) $E_{dc} = 4$ kV/cm. $T = 1.8$ K, whispering gallery mode $m = 128$.

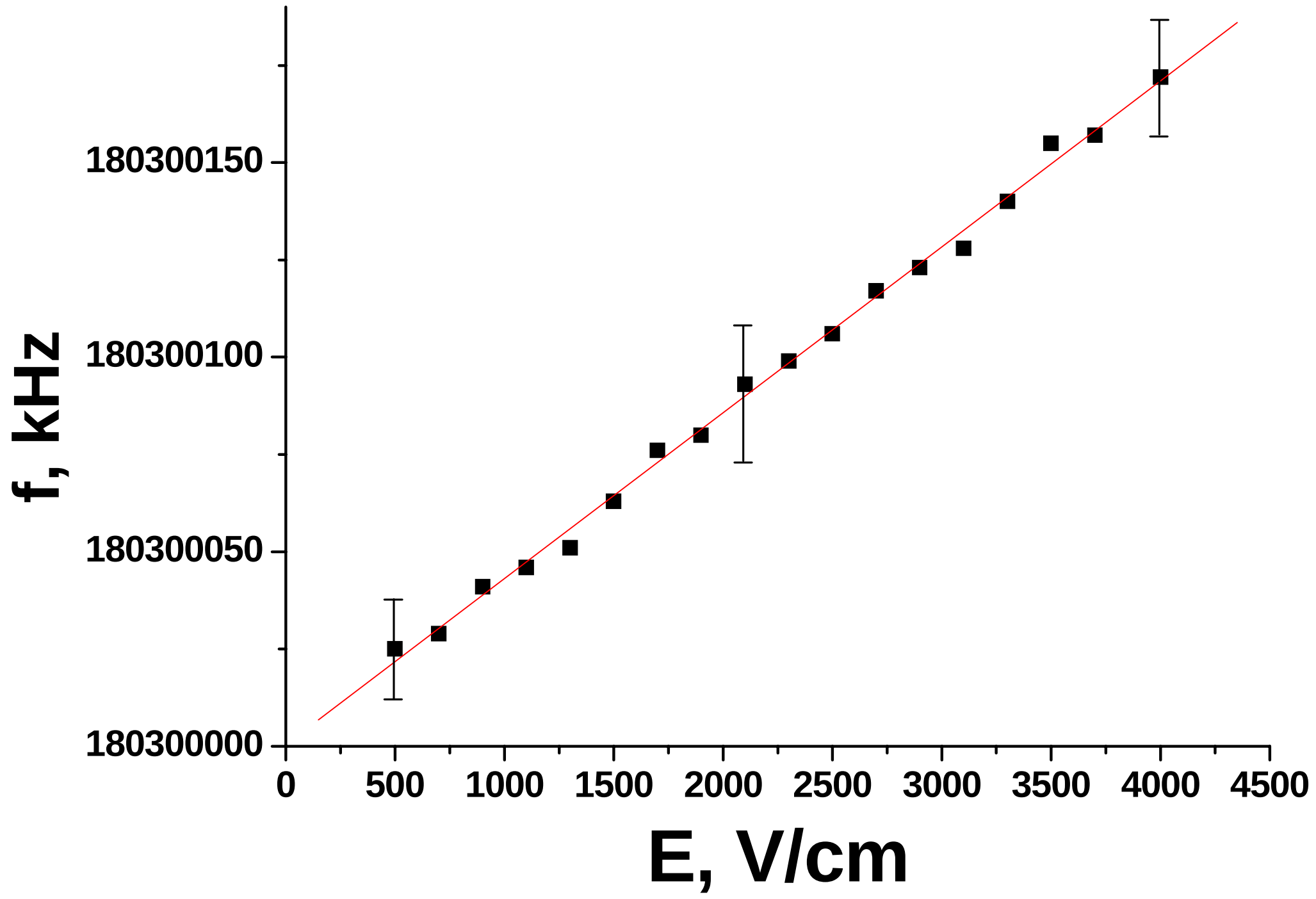

Fig. 5. Frequency shift of the roton resonance line as a function of *dc* electric field. $T = 1.8$ K, $m = 128$. Solid line – approximation by Eq. (1).